\documentclass[preprint,aps,showpacs,nofootinbib,preprintnumbers,amsmath,amssymb]{revtex4-1}
\usepackage{amssymb}
\usepackage{epsfig}
\usepackage{graphicx}% Include figure files
\usepackage{subfigure}
\usepackage{dcolumn}% Align table columns on decimal point
\usepackage{bm}% bold math
\usepackage[usenames ,dvipsnames]{xcolor}
\usepackage{slashed}

\begin{document}

\title{Large $\nu$-$\bar{\nu}$ Oscillations from High-Dimensional Lepton Number Violating Operator}

\author{Chao-Qiang~Geng$^{1,2,3}$\footnote{geng@phys.nthu.edu.tw} and
Da~Huang$^{2,4}$\footnote{dahuang@fuw.edu.pl}
}
 \affiliation{$^{1}$Chongqing University of Posts \& Telecommunications, Chongqing, 400065, China\\
  $^{2}$Department of Physics, National Tsing Hua University, Hsinchu, Taiwan\\
  $^{3}$Physics Division, National Center for Theoretical Sciences, Hsinchu, Taiwan \\
  $^{4}$Faculty of Physics, University of Warsaw, Pasteura 5, 02-093 Warsaw, Poland}

\date{\today}
\begin{abstract}
It is usually believed that the observation of the neutrino-antineutrino (${\nu}$-$\bar{\nu}$) oscillations is almost impossible since the oscillation probabilities are expected to be greatly suppressed by the square of tiny ratio of neutrino masses to energies. Such an argument is applicable to most models for neutrino mass generation based on the Weinberg operator, including the seesaw models. However, in the present paper, we shall give a counterexample to this argument, and show that large $\nu$-$\bar{\nu}$ oscillation probabilities can be obtained in a class of models in which both neutrino masses and neutrinoless double beta ($0\nu\beta\beta$) decays are induced by the high-dimensional lepton number violating operator ${\cal O}_7 = \bar{u}_R l^c_R \bar{L}_L H^*d_R + {\rm H.c.}$  with $u$ and $d$ representing the first two generations of quarks. In particular, we find that the predicted $0\nu\beta\beta$ decay rates have already placed interesting constraints on the $\nu_e \leftrightarrow \bar{\nu}_e$ oscillation. Moreover, we provide an UV-complete model to realize this scenario, in which a dark matter candidate naturally appears due to the new $U(1)_d$ symmetry.
\end{abstract}

%\pacs{ }
%\keywords{ }
\maketitle

\section{Introduction}\label{s1}
It is well established that neutrinos have tiny masses and mixings by observing neutrino oscillations between different flavors~\cite{Anselmann:1992kc,Fukuda:1998mi,Ahmad:2002jz,Ahmad:2002ka,Ahn:2006zza,
Abe:2011sj,An:2012eh}. Such phenomena cannot be interpreted within the Standard Model (SM), giving us one of the strong motivations towards new physics. 

\begin{figure}[th]
\includegraphics[scale=0.8]{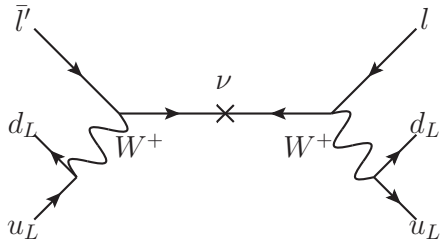}
\caption{Feynman diagram for the conventional modes of neutrino-antineutrino oscillations.}
\label{FignanOs_c}
\end{figure}

In order to generate the neutrino masses in a natural way, most models in the literature, including the 
seesaw~\cite{TypeIseesaw1,TypeIseesaw2,TypeIseesaw3,TypeIseesaw4,TypeIseesaw5,typeIIseesaw1,typeIIseesaw2,typeIIseesaw3,
typeIIseesaw4,typeIIseesaw5,typeIIseesaw6,typeIIseesaw7,Foot:1988aq} and radiative neutrino mass 
generation~\cite{Krauss:2002px,Ma:2006km,Aoki:2008av,Gustafsson:2012vj, Zee:1980ai,Zee:1985id,Babu:1988ki,Chen:2006vn,Chen:2007dc,Jin:2015cla,Geng:2014gua,Geng:2015coa, Geng:2015qha,Geng:2015oga} mechanisms, 
require the  Majorana-type of neutrinos, which implies the lepton number violation (LNV). A traditional smoking gun for 
the LNV is the neutrinoless double beta ($0\nu\beta\beta$) decays~\cite{Pas:2015eia}. However, the LNV effects are usually predicted to be very small in most models based on the conventional dimension-5 Weinberg operator, since the corresponding amplitudes are always induced by the tiny Majorana neutrino masses and thus greatly suppressed. One prominent example is the neutrino-antineutrino ($\nu \to \bar{\nu}$) oscillations~\cite{Schechter:1980gk,Li:1981um,Bernabeu:1982vi, Kayser:1984ge,Langacker:1998pv,deGouvea:2002gf,Xing:2013ty,Xing:2013woa, Zhou:2013eoa,Xing:2014yka,Xing:2014eia} with the relevant Feynman diagram shown in Fig.~\ref{FignanOs_c}.    
It is clear that, compared with the usual (anti)neutrino-(anti)neutrino oscillations, there is an extra $(m_\nu/E_\nu)^2$ suppression in the oscillation probabilities of these $\nu \to \bar{\nu}$ modes, where $m_\nu$ ($E_\nu$) denotes the neutrino mass (energy). For reactor neutrinos with energies of ${\cal O}(1)$~MeV, such a suppression factor would be of ${\cal O}(10^{-16})$, while for accelerator neutrinos with $E_\nu \sim {\cal O}(1)$~GeV, this factor is even of ${\cal O}(10^{-22})$. Therefore, it is almost impossible to observe this LNV phenomenon experimentally in the near future. 

However, in the present paper, we shall show that the restrictions above 
could be waived for
%cannot be applicable to 
a class of models with the neutrino masses generated from the following specific dimension-7 LNV operator~\cite{Babu:2001ex,deGouvea:2007qla, Angel:2012ug,Helo:2015fba,Babu:2010vp,Cai:2014kra}:
\begin{eqnarray}\label{EqO7}
{\cal O}_7 =  \sum_{u,d,l,l^\prime} \frac{C^{ud}_{l l^\prime}}{\Lambda^3}\,\bar{l}_R^c u_R  \bar{d}_R \tilde{H}^T L_L^\prime + {\rm H.c.}\,,
\end{eqnarray}
which is pictorially shown in Fig.~\ref{FigO7}a\footnote{Note that the subscript under ${\cal O}$ denotes the mass dimension of the effective operator throughout the paper. In fact, the operator ${\cal O}_7$ in Eq.~(\ref{EqO7}) is actually ${\cal O}_8$ in the Babu-Leung list~\cite{Babu:2001ex} of the LNV operators up to dimension-11, and has already been throughly studied in Refs.~\cite{Babu:2010vp,Cai:2014kra}. But it has not yet been pointed out the interesting new neutrino-antineutrino modes in this scenario.}. Here, $u_R$, $d_R$, $l_R$ represent the right-handed up-type quarks, down-type quarks and charged leptons, while $L_L$ and $H$ the left-handed lepton and Higgs $SU(2)_L$ doublets with $\tilde{H}\equiv i\sigma^2 H$, respectively. In such a kind of models  with sizable couplings only to the first two generations of quarks, the neutrino-antineutrino oscillation probabilities can be large enough so that such phenomena might be observed with the usual (anti)neutrino sources in the foreseeable future.

The paper is organized as follows. In Sec.~\ref{s2}, we introduce the effective operator ${\cal O}_7$ and 
calculate its contributions to the  neutrino masses and  $0\nu\beta\beta$ decays. We discuss the neutrino-antineutrino oscillations in the present scenario by computing the corresponding oscillation probabilities and $CP$ asymmetries
in Sec.~\ref{s3}. In Sec.~\ref{s4}, we provide a new UV-complete model to realize this scenario in which ${\cal O}_7$ dominates the generation of neutrino masses and LNV effects. Finally, we give the conclusions in Sec.~\ref{s5}.

\section{The Effective Operator, Neutrino Masses, and Neutrinoless Double Beta Decays}\label{s2}
Let us first  consider the neutrino masses generated from the effective operators in Eq.~(\ref{EqO7}).
\begin{figure}[th]
\includegraphics[scale=0.8]{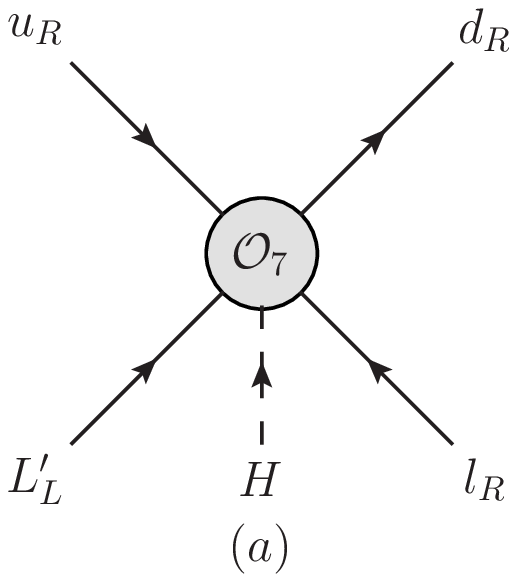}
\includegraphics[scale=0.8]{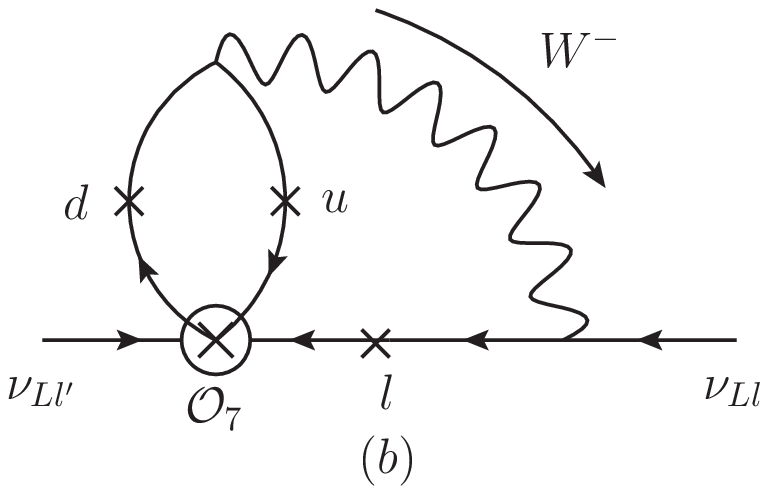}
\caption{Feynman diagrams for (a) effective operator ${\cal O}_7$ and (b) neutrino masses induced by ${\cal O}_7$.  }
\label{FigO7}
\end{figure}
Note that ${\cal O}_7$ breaks the lepton number under the $U(1)$ symmetry by two units, which is the necessary condition to 
produce the Majorana neutrino masses. After the spontaneous electroweak symmetry breaking, ${\cal O}_7$ can induce the following set of dimension-6 effective operators:
\begin{eqnarray}\label{O6Eff}
\tilde{\cal O}_6 = \sum_{u,d,l,l^\prime} \frac{C^{ud}_{l l^\prime} v_0}{\sqrt{2}\Lambda^3} \, \bar{l}_R^c u_R \bar{d}_R \nu_{L l^\prime} + {\rm H.c.}\,,
\end{eqnarray}
where we have redefined the Wilson coefficients $C^{ud}_{l l^\prime}$ into the basis 
with $u_R$, $d_R$ and $l_R$ being the right-handed particles of the mass eigenstates while $\nu_{Ll}$  
still the flavor eigenstates. With $\tilde{\cal O}_6$, it is straightforward to draw the two-loop Feynman diagrams as depicted 
in Fig.~\ref{FigO7}b, assumed to be the dominant contributions to the neutrino masses. 
 The explicit calculation gives the neutrino mass matrix as follows:
\begin{eqnarray}\label{MnMass_o}
(m_\nu)_{ll^\prime} &=&  \frac{g_2^2 v_0}{2\sqrt{2}\Lambda^3} \sum_{u,d} m_u m_d V_{ud} (m_l C^{ud}_{l l^\prime} + m_{l^\prime} C_{l^\prime l}^{ud}) {\cal I}(m_W^2, m_u^2, m_d^2) \nonumber\\
&\approx & \frac{1}{(16\pi^2)^2} \frac{\sqrt{2}}{v_0 \Lambda}\sum_{u,d} m_u m_d V_{ud}(m_l C^{ud}_{l l^\prime} + m_{l^\prime} C_{l^\prime l}^{ud})\,, 
\end{eqnarray}
where $m_{u,d}$ denotes the three generations of up- and down-type quark masses, and $V_{ud}$ the corresponding CKM matrix elements. Since the loop integrals ${\cal I}(m_W^2, m_u^2, m_d^2)$ are quadratic divergent, 
we have estimated that ${\cal I} \sim \Lambda^2/(16\pi^2 m_W)^2 = 4 \Lambda^2/(16\pi^2 g_2 v_0)^2$ in the second equality. If we assume all Wilson coefficients to be of ${\cal O}(1)$, the neutrino masses are dominated  by the top-bottom terms
\begin{eqnarray}\label{MnMass_tb}
(m_\nu)_{l l^\prime} &\approx & \frac{1}{(16\pi^2)^2} \frac{\sqrt{2}}{v_0\Lambda} m_t m_b V_{tb} (m_l C^{tb}_{l l^\prime} + m_{l^\prime} C_{l^\prime l}^{tb})\,
\end{eqnarray}
by considering the quark mass hierarchies of $m_t \gg m_c \gg m_u$ and $m_b \gg m_s \gg m_d$. It follows that, by taking the $\tau$ mass as the typical lepton mass scale, the measured neutrino masses $m_\nu \sim {\cal O}(5 \times 10^{-2}~{\rm eV})$ dictate $\Lambda \sim 6\times 10^3$~TeV~\cite{Babu:2001ex,deGouvea:2007qla,Angel:2012ug}, which is too large to have any observable LNV effects at low energies. However, if the couplings in Eq.~(\ref{EqO7}) with the third-generation quarks are greatly suppressed by, for example, some flavor symmetries, the second-generation quark couplings would give the dominant contribution to neutrino masses
\begin{eqnarray}\label{MnMass}
(m_\nu)_{ll^\prime} &\approx & \frac{1}{(16\pi^2)^2} \frac{\sqrt{2}}{v_0\Lambda} m_c m_s V_{cs} (m_l C^{cs}_{l l^\prime} + m_{l^\prime} C_{l^\prime l}^{cs})\,,
\end{eqnarray}    
leading to the UV cutoff to be $\Lambda \sim 1$~TeV. Such a low UV cutoff makes it possible to detect sizeable LNV effects, so we shall take it in the following discussions.

Note also that the predicted neutrino mass matrix elements are proportional to the charged lepton masses. By taking into account the charged lepton mass hierarchy: $m_\tau > m_\mu \gg m_e$, it is generically expected that the component $(m_\nu)_{ee}$ should be much smaller than other elements in the neutrino mass matrix. Effectively, we can take the approximation $(m_\nu)_{ee} \approx 0$, which reduces to the well-studied texture-zero matrix. According to Refs.~\cite{PDG,Pascoli:2001by,Geng:2015qha,Geng:2015oga}, only the normal ordering can fit the current neutrino oscillation data, which can be regarded as one of predictions of this scenario. 
Moreover, the nearly vanishing $(m_\nu)_{ee}$ even restricts the lightest neutrino mass to be located within the range $0.001\,{\rm eV} \lesssim m_0 \lesssim 0.01\,{\rm eV}$, and the Majorana phase $\alpha_{21}$ in the standard parametrization of the Pontecorvo-Maki-Nakagawa-Sakata (PMNS) mixing matrix~\cite{Pontecorvo:1957cp,Maki:1962mu} to be $0.8\pi \lesssim \alpha_{21}\lesssim 1.2\pi$~\cite{Geng:2015qha,Geng:2015oga}. 

\begin{figure}[th]
\includegraphics[scale=0.7]{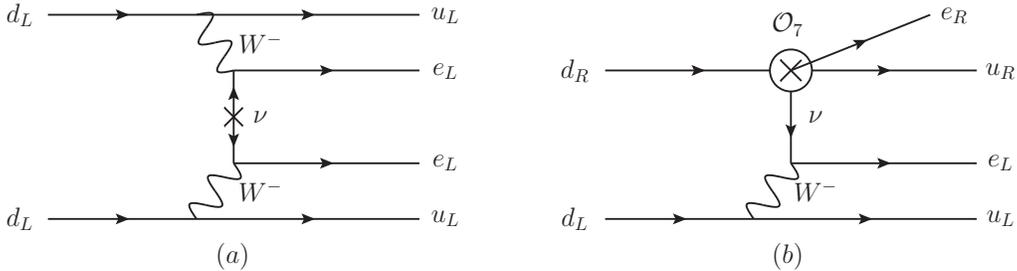}
\caption{Feynman diagrams for neutrinoless double beta decay: (a) Conventional mode; (b) New mode induced by ${\cal O}_7$. }
\label{Fig0nbb}
\end{figure}

%\section{Neutrinoless Double Beta Decay with ${\cal O}_7$}\label{s3}
%We have already mentioned that the effective operators ${\cal O}_7$ break the lepton number $U(1)$ symmetry by two units. T
It is known that the  smoking gun for the LNV is the neutrinoless double beta decay processes. For Majorana neutrinos, there always exists the traditional long-range channel shown in Fig.~\ref{Fig0nbb}a, in which the LNV is induced by the insertion of the neutrino mass $(m_\nu)_{ee}$ so as to flip the chirality of the internal neutrinos. Therefore, the detection of the $0\nu\beta\beta$ processes could help determine $|(m_\nu)_{ee}|$. Nevertheless, it  has already been noted that in some types of neutrino models~\cite{Chen:2006vn,Chen:2007dc,Gustafsson:2012vj,Jin:2015cla,Geng:2015qha,Helo:2015fba, delAguila:2011gr,delAguila:2012nu,Gustafsson:2014vpa,King:2014uha}, Fig.~\ref{Fig0nbb}a does not give the main contribution, whereas some other modes would dominate, as just the case for the present scenario shown in Fig.~\ref{Fig0nbb}b.

%The new Feynman diagrams for the $0\nu\beta\beta$ decays are shown i
 In Fig.~\ref{Fig0nbb}b,  
 the LNV occurs at the vertices due to the insertions of $\tilde{\cal O}_6$. In order to apply the formalism in Refs.~\cite{Doi:1985dx, Muto:1989cd,Hirsch:1996qw}, we need first to transform the scalar-scalar interactions in Eq.~(\ref{O6Eff}) into the desired form in terms of vector-vector ones:
\begin{eqnarray}\label{O6_Tr}
\tilde{\cal O}_6 = \frac{v_0}{8\sqrt{2}\Lambda^3}\sum_{u,d,l,l^\prime} C^{ud\,*}_{ll^\prime} \bar{u} \gamma_\mu (1+\gamma_5) d \,\bar{l} \gamma^\mu (1+\gamma_5)\nu^c_{l^\prime}  + {\rm H.c.}\,,
\end{eqnarray}
where we have used the Fierz identity and made the charge conjugations of the currents. We can then extract the relevant part for the $0\nu\beta\beta$ decays
\begin{eqnarray}
{\cal O}^{ud}_{ee} = \lambda \frac{G_F}{\sqrt{2}} \bar{u} \gamma^\mu (1+\gamma_5) d \bar{e} \gamma_\mu (1+\gamma_5) \nu_e^c\,,
\end{eqnarray}
since the quarks and charged leptons involved in the process are $u$, $d$ and $e$, respectively. Note that in the above expression, we have defined the interaction coupling as
\begin{eqnarray}
\lambda = \frac{C_{ee}^{ud\,*} v_0}{8G_F \Lambda^3}\,,
\end{eqnarray}
according to the conventional formalism in Refs.~\cite{Doi:1985dx, Muto:1989cd,Hirsch:1996qw}.

By taking into account both decay channels in Fig.~\ref{Fig0nbb}, we obtain the general formula for the half lifetime $T_{1/2}^{0\nu\beta\beta}$ as follows~\cite{Muto:1989cd}:
\begin{eqnarray}
\label{eq8}
\left[ T^{0\nu\beta\beta}_{1/2} \right]^{-1} = C_{mm} \left( \frac{(m_{\nu})_{ee}}{m_e} \right)^2  + C_{m\lambda} \left(\frac{(m_\nu)_{ee}}{m_e} \right)\lambda + C_{\lambda\lambda}\lambda^2\,,
\end{eqnarray}
where $C_{mm}$, $C_{m\lambda}$, and $C_{\lambda\lambda}$ can be determined by the corresponding phase space integrations and nuclear matrix elements. By using the numerical values of $C$ listed in  Table 2 of Refs.~\cite{Muto:1989cd}, 
 we find that the first two terms in Eq.~(\ref{eq8}) are much smaller than the last one 
%the first two terms since the latter is suppressed by 
due to the suppressions of the nearly vanishing neutrino mass $(m_\nu)_{ee}$. 
As a result, it is concluded that the new long-range mode induced by ${\cal O}_7$ gives the dominant contribution to the $0\nu\beta\beta$ decays in the present scenario.

By comparing the current experimental limits on the $0\nu\beta\beta$ half-life of different nuclei~\cite{Agostini:2013mzu,KamLANDZen,Gando:2012zm,Argyriades:2008pr, Arnaboldi:2008ds,Arnold:2005rz,Barabash:2010bd}, 
we can actually give the strong constraints to the Wilson coefficient $ C^{ud}_{ee}$ by assuming $\Lambda \sim 1$~TeV. 
The numerical limits are collected in Table~\ref{Tab_0nbb}, from which it is seen that the strongest constraint is given by the target $^{136}$Xe~\cite{KamLANDZen,Gando:2012zm} with  $|C^{ud}_{ee}|<1.9\times 10^{-4} $. Finally, we should mention that, contrary to the general  expectation, such a stringent upper bound on $C^{ud}_{ee}$ does not place any constraint to the neutrino mass element $(m_\nu)_{ee}$, since the neutrino mass formula in Eq.~(\ref{MnMass}) depends on the different Wilson coefficients  $C^{cs}_{ll^\prime}$ related to the second-generation quarks.

\begin{table}[ht]
\caption{Constraints on $\lambda$ and $|C_{ee}^{ud}|$ from $0\nu\beta\beta$ for different target nuclei by assuming $\Lambda = 1$~TeV. }
\begin{tabular}{l|c|c|c|cl}
\hline
&\;\;$T_{\rm exp}(10^{25}{\rm yr})$\;\;&$C_{\lambda\lambda}\,({\rm yr^{-1}})$ & $\lambda$ & $|C^{ud}_{ee}|$\\
\hline
GERDA-1($^{76}$Ge)~\cite{Agostini:2013mzu}&2.1&$1.36\times10^{-13}$&$5.9\times10^{-7}$&$2.2\times10^{-4}$\\
KamLAND-Zen($^{136}$Xe)~\cite{KamLANDZen,Gando:2012zm}&1.9&$2.04\times10^{-13}$&$5.1\times10^{-7}$&$1.9\times10^{-4}$\\
NEMO-3($^{150}$Nd)~\cite{Argyriades:2008pr}&0.0018&$2.68\times10^{-11}$&$1.4\times10^{-6}$&$5.5\times10^{-4}$\\
CUORICINO($^{130}$Te)~\cite{Arnaboldi:2008ds}&0.3&$1.05\times10^{-12}$&$5.6\times10^{-7}$&$2.1\times10^{-4}$\\
NEMO-3($^{82}$Se)~\cite{Arnold:2005rz,Barabash:2010bd}&0.036&$1.01\times10^{-12}$&$1.7\times10^{-6}$&$6.3\times10^{-4}$\\
NEMO-3($^{100}$Mo)~\cite{Barabash:2010bd}&0.11&$1.05\times10^{-12}$&$9.3\times10^{-7}$&$3.5\times10^{-4}$\\
\hline
\end{tabular}
\label{Tab_0nbb}
\end{table}

More recently, there is some interest in the estimation of the sensitivity of the $\mu^- - e^+$ conversion in nuclei in the literature~\cite{Geib:2016atx,Berryman:2016slh,Geib:2016daa}, given that the the sensitivity of this mode is expected to be increased greatly in the near future due to the tremendous experimental improvement in the similar $\mu^- - e^-$ conversion mode. Nowadays, the most stringent limit on this channel was set by the SINDRUM II Collaboration in 1998, with the $90\%$ C.L. upper bound on the rate as follows~\cite{Kaulard:1998rb}:
\begin{eqnarray}\label{BmeLNV}
R^{\rm Ti}_{\mu^- e^+} \equiv \frac{\Gamma(\mu^- + {\rm Ti} \to e^+ + {\rm Ca})}{\Gamma(\mu^- + {\rm Ti} \to \nu_\mu + {\rm Sc})} < 1.7\times 10^{-12}\,,
\end{eqnarray}
which was obtained by assuming that the process occurs by the coherent scattering to the ground state of calcium. With the forthcoming next-generation $\mu^- - e^-$ conversion experiments such as Mu2e~\cite{Mu2e} at Fermilab and COMET~\cite{COMET} at J-PARC in Japan, the sensitivities to the $\mu^- - e^+$ conversion are expected to reach $R^{\rm Al}_{\mu^- e^+} \sim 10^{-16}$ and $R^{\rm Al}_{\mu^- e^+} \sim 10^{-14}$~\cite{Berryman:2016slh}, respectively. Following Ref.~\cite{Berryman:2016slh}, we can estimate the $\mu^- - e^+$ conversion rate in our scenario to be
\begin{eqnarray}\label{RmeLNV}
\Gamma(\mu^- - e^+) \sim \left|\frac{(C^{ud}_{\mu e}+ C^{ud}_{e\mu}) v_0}{8\sqrt{2}\Lambda^3} \right|^2 \left(\frac{G_F}{\sqrt{2}}\right)^2 \left(\frac{Q^8}{q^2}\right)|\psi_{100}(0)|^2\,,
\end{eqnarray}
where we have neglected the two-loop contribution to the $\mu^-$-$e^+$ conversion rate, which is subdominant in our considered cutoff scale $\Lambda\sim 1$~TeV. The first factor in Eq.~(\ref{RmeLNV}) comes from the coefficient before the LNV operator according to our conventions. The factor $G_F/\sqrt{2}$ is contributed by the $W$-boson propagator and its couplings, while $1/q^2$ is the dominant contribution from the neutrino propagator and estimated to be of ${\cal O}(1/(100~{\rm MeV})^2)$, which is the typical distance between nucleons in a nuclei. $|\psi_{100}(0)|^2$ is the 1s ground state probability density function of the muon in the captured atom, which has a mass dimension of 3. Finally, all the other quantities related to the phase-space and nuclear matrix elements are characterized by the energy scale $Q$, and the power of $Q$ is determined by the requirement that the final expression has the mass dimension of a decay rate. By fitting the known nuclear matrix element of titanium for the long-range light neutrino exchange, the scale $Q$ is estimated to be 15.6~MeV~\cite{Berryman:2016slh}. By a similar argument, the muon capture rate can be approximated as
\begin{eqnarray}\label{Rmc}
\Gamma_{\mu c} \sim \left(\frac{G_F}{\sqrt{2}}\right)^2 Q^2 |\psi_{100}(0)|^2\,.
\end{eqnarray} 
Thus, by taking the ratio between Eqs.~(\ref{RmeLNV}) and (\ref{Rmc}), we can obtain
\begin{eqnarray}
R_{\mu^- e^+} \sim \left|\frac{(C^{ud}_{\mu e}+ C^{ud}_{e\mu}) v_0}{8\sqrt{2}\Lambda^3} \right|^2  \left(\frac{Q^6}{q^2}\right)\,.
\end{eqnarray}
If we take $\Lambda \sim 1$~TeV and the Wilson coefficient $C^{ud}_{\mu e,~ e\mu} \sim 1$, this ratio is of order of $10^{-24}$, which is too small compared with the current bound in Eq.~(\ref{BmeLNV}) and the sensitivity of next-generation experiments.

For other LNV channels, such as rare meson decays $K^\pm \to \pi^\mp \mu^\pm \mu^\pm$~\cite{Massri:2016dff}, $D^+ \to K^- e^+ \mu^+$~\cite{Lees:2011hb} and rare tau decays $\tau^- \to e^+ \pi^- \pi^-$~\cite{Miyazaki:2012mx}, the sensitivities are even lower, so that we expect that they do not give rise to strong constraints to the present scenario. Therefore, we do not consider them in our following discussion.

\section{Neutrino-Antineutrino Oscillations via ${\cal O}_7$}\label{s3}
\begin{figure}[th]
\includegraphics[scale=0.8]{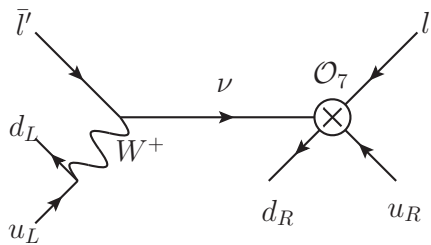}
\caption{Feynman diagrams for neutrino-antineutrino oscillations. }
\label{FignanOs}
\end{figure}

Now we consider the neutrino-antineutrino oscillations when the effective operator ${\cal O}_7$ gives rise to the measured Majorana neutrino masses. Besides of the conventional mechanism with the neutrino mass insertion shown in Fig.~\ref{FignanOs_c}, ${\cal O}_7$ can generate the new contributions to the phenomena of $\nu_l \to \bar{\nu}_{l^\prime}$, with the Feynman diagrams presented in Fig.~\ref{FignanOs}. Note that the LNV only occurs at each of two vertices in these new amplitudes, so that it is possible to
induce large oscillation probabilities by avoiding the huge suppression from the mass insertions.
Since the neutrino mass eigenstates propagate in space, we need first to transform the neutrino fields in Eq.~(\ref{O6_Tr}) into their mass states:
\begin{eqnarray}
\tilde{\cal O}_6 = -\frac{v_0}{8\sqrt{2}\Lambda^3}\sum_{u,d,l,j} \tilde{C}^{ud}_{lj} \bar{d} \gamma_\mu (1+\gamma_5) u \,\bar{l^c} \gamma^\mu (1-\gamma_5)\nu_{j}  + {\rm H.c.}\,,
\end{eqnarray}
in which the relation $\nu_l = \sum_j U_{lj}\nu_j$ has been used to define
\begin{eqnarray}
\tilde{C}^{ud}_{lj} = \sum_{l^\prime} C^{ud}_{ll^\prime} U_{l^\prime j}\,,
\end{eqnarray}
where $U$ represents the PMNS matrix. The amplitude for this new contribution to $\nu_l \to \bar{\nu}_{l^\prime}$ is thus given by
\begin{eqnarray}\label{AmpNO}
i {\cal M}(\nu \to \bar{\nu}_{l^\prime}) &=& {\cal K}_{\nu \bar{\nu}}\left(\frac{v_0 G_F}{16\Lambda^3}\right) \sum_j (\tilde{C}^{ud}_{lj}U^*_{l^\prime j} + U_{lj}^* \tilde{C}^{ud}_{l^\prime j})D_j\nonumber\\
&\approx& {\cal K}_{\nu \bar{\nu}}\left(\frac{v_0 G_F}{16\Lambda^3}\right) \sum_j \Gamma^{l^\prime l}_j e^{-\frac{iL}{2E}m_j^2}\,,
\end{eqnarray}
where we have approximated the neutrino wave-functions as
\begin{equation}
D_j = e^{-i (E_j T - p_j L)} \approx e^{-iLm_j^2/2E}
\end{equation}
by assuming neutrinos propagate nearly at the speed of light so that their momenta can be estimated as 
$p_j \approx E_j - (m_j^2/2E_j)$. For simplicity, we have also defined $\Gamma^{l^\prime l}_j$ as the coefficients involving the products of $C^{ud}_{lj}$ and $U_{lj}$, and used ${\cal K}_{\nu \bar{\nu}}$ to denote the relevant nuclear form  
and kinematic factors, which are usually chosen to be real. 
Consequently, 
%By taking the square of the amplitudes, we can yield 
the probabilities for the neutrino-antineutrino oscillations are given by
\begin{eqnarray}\label{OsP}
P(\nu_l \to \bar{\nu}_{l^\prime}) &=& {\cal K}_{\nu \bar{\nu}}^2 \left(\frac{v_0 G_F}{16\Lambda^3}\right)^2 \sum_{j,k} \Gamma^{l^\prime l}_{j} \Gamma^{l^\prime l\,*}_k e^{-i\varphi_{jk}} \nonumber\\
&=& {\cal K}_{\nu \bar{\nu}}^2 \left(\frac{v_0 G_F}{16\Lambda^3}\right)^2 \{\sum_j |\Gamma^{l^\prime l}_j|^2 \nonumber\\
&&  + 2\sum_{j>k} [{\rm Re}(\Gamma^{l^\prime l}_j \Gamma^{l^\prime l\,*}_{k})\cos\varphi_{jk} + {\rm Im}(\Gamma^{l^\prime l}_j \Gamma^{l^\prime l\,*}_{k})\sin\varphi_{jk}] \} \,,
\end{eqnarray}
where we define the strong phases as $\varphi_{jk} = L(m_j^2-m_k^2)/(2E)$. In general, the matrix $\Gamma_j^{l^\prime l}$ is complex, so that it is expected that $CP$ violation can be observed in the $\nu_l \to \bar{\nu}_{l^\prime}$ oscillations. The $CP$ conjugate process is $\bar{\nu}_l \to \nu_{l^\prime}$, for which the oscillation probability can be obtained as follows
\begin{eqnarray}\label{OsPBar}
P(\bar{\nu}_l \to \nu_{l^\prime}) &=& {\cal K}_{\nu \bar{\nu}}^2 \left(\frac{v_0 G_F}{16\Lambda^3}\right)^2 \{\sum_j |\Gamma^{l^\prime l}_j|^2 \nonumber\\
&&  + 2\sum_{j>k} [{\rm Re}(\Gamma^{l^\prime l}_j \Gamma^{l^\prime l\,*}_{k})\cos\varphi_{jk} - {\rm Im}(\Gamma^{l^\prime l}_j \Gamma^{l^\prime l\,*}_{k})\sin\varphi_{jk}] \} \,.
\end{eqnarray}
By taking the difference of the above two formula, we can obtain the $CP$ asymmetry  
for the neutrino-anti-neutrino oscillation channels between flavors $l$ and $l^\prime$ as follows:
\begin{eqnarray}
{\cal A}_{CP}^{l^\prime l} &\equiv & P(\nu_l \to \bar{\nu}_{l^\prime}) - P(\bar{\nu}_l \to \nu_{l^\prime})\nonumber\\
&=& 4 {\cal K}_{\nu \bar{\nu}}^2 \left(\frac{v_0 G_F}{16\Lambda^3}\right)^2 \sum_{j>k} {\rm Im}(\Gamma^{l^\prime l}_j \Gamma^{l^\prime l\,*}_{k})\sin\varphi_{jk} \,.
\end{eqnarray}
We now estimate the typical size of the probabilities for $\nu_l \to \bar{\nu}_{l^\prime} $ by taking the ratio of Eq.~(\ref{OsP}) with the corresponding neutrino-neutrino oscillations of the same flavor dependences:
\begin{eqnarray}
\frac{P(\nu_l \to \bar{\nu}_{l^\prime})}{P(\nu_l \to \nu_{l^\prime})} &\approx & \left(\frac{{\cal K}_{\nu\bar{\nu}}}{{\cal K}_{\nu}}\right)^2 \left(\frac{v_0 G_F/16\Lambda^3}{G_F^2/2}\right)^2 |C^{ud}_{l^\prime l}|^2\nonumber\\
&=& \left(\frac{{\cal K}_{\nu\bar{\nu}}}{{\cal K}_{\nu}}\right)^2 \left(\frac{v_0^3}{4 \Lambda^3}\right)^2 |C^{ud}_{l^\prime l}|^2\,,
\end{eqnarray}
where ${\cal K}_\nu$ denotes the form factors for the conventional neutrino-neutrino oscillations,
 assuming to be ${\cal K}_{\nu}\approx {\cal K}_{\nu\bar{\nu}}$. 
If $\Lambda \sim 1$~TeV and $C^{ud}_{l^\prime l} \sim 1$ from neutrino mass calculations, the neutrino-antineutrino oscillations can be only mildly suppressed with a factor of ${\cal O}(10^{-6})$ compared with the neutrino-neutrino counterparts. However, as shown in the previous section, the $0\nu\beta\beta$ decay experiments have already presented strong limits on the elements $C^{ud}_{ee}<1.9\times 10^{-4}$. Thus, we expect that the $\nu_e \leftrightarrow \bar{\nu}_e$ channel should be much smaller than other channels. Except for $\nu_e \leftrightarrow \bar{\nu}_e$, other modes are not  much constrained currently, so that their amplitudes can be large, and provide interesting signatures for this neutrino mass generation mechanism. Especially, opposed to the conventional channels in Fig.~\ref{FignanOs_c}, the oscillation probabilities induced by ${\cal O}_7$ do not depend much on the neutrino energies. Thus, it opens the possibility to use conventional (anti)neutrino sources to detect such phenomena, such as the reactor and accelerator neutrinos. 

%Furthermore, the dependence on the strong phases $\varphi_{jk}$ in the $\nu \to \bar{\nu}$ channels is the same as those of the neutrino oscillations, so that the same facilities can be applied to detect them.

\section{A Model Realizing ${\cal O}_7$ with Dark Matter }\label{s4}
\begin{figure}[th]
\includegraphics[scale=0.7]{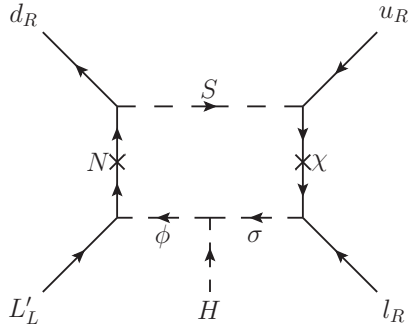}
\caption{Feynman diagrams for generating ${\cal O}_7$ in the model. }
\label{FigModel}
\end{figure}

In this section, we present a new UV-complete model to realize ${\cal O}_7$ as the leading-order LNV effects. The new  fields with their charge assignments are listed in Table~\ref{Tab_model}, in which $\sigma$, $S$, and $\phi$ are complex scalars, while $\chi$ and $N$ are vector-like fermions. We also impose a new $U(1)_d$ symmetry, under which all SM fields are neutral and only new particles are charged.
\begin{table}[ht]
\caption{Charge assignments of new fields in the dark sector}
\begin{tabular}{c|c|c|c|c}
\hline
 & $SU(3)_C$ & $SU(2)_L$ & $U(1)_Y$ & $U(1)_d$\\
\hline
$\sigma_i$ & {\bf 1} & {\bf 1} & 0 & 1 \\
$S$ & {\bf 3} & {\bf 1} & $-2/3$ & $-1$ \\
$\phi_i$ & {\bf 1} & {\bf 2} & $-1$ & $ -1 $\\
$ \chi_{L,R} $ & {\bf 1} & {\bf 1} & 2 & 1\\
$ N_{L,R} $ & {\bf 1} & {\bf 1} & 0 & 1 \\
\hline
\end{tabular}
\label{Tab_model}
\end{table}
The relevant Lagrangian involving the new fields is given by
\begin{eqnarray}\label{Lag_d}
-{\cal L}_d &=&  f_{li} \phi_i^\dagger \bar{N}_R L_L + g_{li} \bar{l}^c_R \chi_R \sigma_i^\dagger + h_u S^\dagger
\bar{\chi}_L u_R + k_d \bar{d}_R N_L S\nonumber \\
& & +  \mu_i \sigma_i \tilde{H}\phi_i + m_\chi \bar{\chi}_R\chi_L + m_N \bar{N}_L N_R + {\rm H.c.} + V(\sigma, S, \phi, H)\,,
\end{eqnarray}
where $V(\sigma_i, \phi_i, S, H)$ represents the scalar potential and is assumed to be stable so that the $U(1)_d$ symmetry keeps 
to be exact. In order to be consistent with our previous effective operator analysis, we require $h_u\sim h_c \gg h_t$ and $h_d \sim h_s \gg h_b$ so that the couplings to the third-generation quarks are suppressed.
 Since the new fermion spectrum is vector-like, there is no gauge anomaly associated with SM gauge groups. 
 Note that the lepton number $U(1)_L$ symmetry is explicitly broken by the Lagrangian in Eq.~(\ref{Lag_d}) 
  only when $\mu_i$, $m_\chi$, $m_N$ and at least one of the products $f_{li} g_{l^\prime i} h_u k_d$ are nonzero simultaneously. 
As a consequence, we can generate the effective operator ${\cal O}_7$ by the one-loop Feynman diagram as shown in Fig.~\ref{FigModel}, while the conventional Weinberg operator is only induced at  higher loops. 
Therefore, ${\cal O}_7$ dominates the Majorana neutrino mass generations and the low-energy LNV processes, which have already been discussed previously.  In particular, by replacing the blob in Fig.~\ref{FigO7}b with the 1-loop box diagram in Fig.~\ref{FigModel}, the Majorana neutrino masses are generated at the three-loop level.
We remark that  to generate a realistic neutrino mass matrix with three different nonzero eigenvalues, 
at least three $\sigma$'s and $\phi$'s are needed,  which are labelled by the subscript $i$ in Eq.~(\ref{Lag_d}).  

 It is quite useful to express the cutoff scale $\Lambda$ and the Wilson coefficients $C^{ud}_{ll^\prime}$ in ${\cal O}_7$ in terms of the parameters in the present model. Direct computations of Fig.~\ref{FigModel} give the following relation:
\begin{eqnarray}\label{CllM}
\frac{C_{ll^\prime}^{ud}}{\Lambda^3} = \frac{1}{16\pi^2} \frac{m_N m_\chi}{m_S^6} h_u k_d \sum_i \mu_i f_{l^\prime i} g_{l i}{\cal I}_1^i\,,
\end{eqnarray} 
where ${\cal I}_1^i$ is the 1-loop integral involving the fields $\phi_i$ and $\sigma_i$. We  have also assumed that the leptoquark mass $m_S$ is the largest mass scale in the loop, so that we can extract correct powers of $m_S$ to make the 1-loop integral ${\cal I}_1^i$ dimensionless and of ${\cal O}(1)$. If we further identify the cutoff scale in Eq.~(\ref{EqO7}) as $\Lambda \equiv m_S$, then the Wilson coefficients correspond to
\begin{eqnarray}
{C_{ll^\prime}^{ud}} = \frac{1}{16\pi^2}  h_u k_d \sum_i \frac{m_N m_\chi \mu_i}{m_S^3} f_{l^\prime i} g_{l i}{\cal I}_1^i\,.
\end{eqnarray}
Such an identification can be further justified by the computation of the three-loop neutrino masses,  given by
\begin{eqnarray}\label{MnM}
(m_\nu)_{ll^\prime} = \frac{1}{(16\pi^2)^3} \frac{g_2^2 v_0 m_N m_\chi}{\sqrt{2} m_W^2 m_S^4} \sum_i \mu_i(g_{li}f_{l^\prime i}  m_l +g_{l^\prime i} f_{l i}m_{l^\prime}) \sum_{u,d} m_u m_d h_u k_d {\cal I}_3^{i,u,d}\,,
\end{eqnarray}
where ${\cal I}_3^{i,u,d}$ are the dimensionless ${\cal O}(1)$ three-loop integrals. If all of $\sigma_i$ and $\phi_i$ have the same masses,  the integrals ${\cal I}_1^i$ and ${\cal I}_3^{i,u,d}$ can take the universal form without the dependence of the index $i$. In this case, the neutrino mass matrix can be reduced to the form as Eq.~(\ref{MnMass_o})  if we take ${\cal I}(m_W^2, m_u^2, m_d^2)\sim {\cal I}_3^{i,u,d}/{\cal I}_1^i$.

 The present model suffers from stringent constraints from the lepton flavor violation (LFV) processes, like $l\to l^\prime \gamma$~\cite{Adam:2013mnn,Aubert:2009ag}, the $\mu^- - e^-$ conversions in nuclei~\cite{Bertl:2006up,Badertscher:1980bt,Dohmen:1993mp,Honecker:1996zf}, and three-body LFV decays $l \to l_1 l_2 \bar{l}_3$~\cite{Bellgardt:1987du,Hayasaka:2010np}, which might spoil our previous arguments based on the effective operator ${\cal O}_7$. But the LNV observables usually involve different coupling dependences from the LFV observables, so it is easy to evade such LFV limits by some level of tuning of model parameters. For example, the process of $\mu^\pm \to e^\pm\gamma$ is among the most sensitive LFV probes since it usually constrains the model most stringently. In our model, there are two kinds of one-loop diagrams contributing to this process, the ones with $\phi^\pm$-$N$ loops and those with $\sigma$-$\chi^\pm$ loops. In order to simplify our discussion, we work with the assumption that the $\phi^\pm$-$N$ loops always dominate over the $\sigma$-$\chi^\pm$ ones. As a result, the main contribution to $\mu^\mp \to e^\mp \gamma$ is
\begin{eqnarray}
{\cal B}(\mu^\mp \to e^\mp ) = \frac{3\alpha}{64\pi G_F^2 } \left|\sum_i  \frac{f_{\mu i} f_{e_i}^*  K({m_N^2}/{m_{\phi_i}^2})}{m_{\phi_i}^2}\right|^2\,,
\end{eqnarray}
where we have defined the loop integral
\begin{equation}
K(z) \equiv \frac{2z^3+3z^2-6z+1-6z^2\log z}{6(1-z)^4}\,.
\end{equation}
By comparing with the upper bound ${\cal B}(\mu^+ \to e^+ \gamma)< 5.7\times 10^{-13}$ from the MEG Collaboration~\cite{Adam:2013mnn} and assuming all of the $\phi$'s have the same mass of order of 1~TeV, we can obtain the following constraint on $f_{\mu i,e i}$:
\begin{eqnarray}\label{Cmeg}
\sum_i f_{\mu i} f^*_{e i} < 5.8\times 10^{-3}\,.
\end{eqnarray} 
With essentially the same argument, the $\sigma$-$\chi^\pm$ loops for  $\mu^\mp \to e^\mp \gamma$ can constrain the combination $\sum_i g_{\mu i} g_{ei}^*$ to a similar order. Furthermore,
we can obtain a slightly less stringent constraint with the limit on the $\mu^- - e^-$ conversion in gold nuclei~\cite{Bertl:2006up}.

In face of the strong limits from the LFV processes, one may worry about their constraints on the neutrino masses and proposed neutrino-antineutrino oscillations, especially for the $\nu_\mu \leftrightarrow \bar{\nu}_e$ channel. However, the mismatch of the coupling dependences in LNV and LFV processes makes the advertised LNV phenomena compatible with these LFV constraints. For instance, if we take $f_{\mu i}$ and $g_{e i}$ to be of ${\cal O}(10^{-3})$ while $f_{e i}$ and $g_{\mu i}$ of ${\cal O}(1)$, the LNV limits are obviously satisfied. In this case, $C^{ud}_{e \mu} \sim O(10^{-6})$ but $C^{ud}_{\mu e}\sim {\cal O}(1)$. However, note that the LNV observables, such as the Majorana neutrino masses in Eq.~(\ref{MnMass_o}) and the amplitude for the $\nu_\mu \leftrightarrow \bar{\nu}_e $ oscillations in Eq.~(\ref{AmpNO}), only rely on the summation of the $C_{\mu e, e\mu}^{ud}$ which are symmetric in the indices $e$ and $\mu$. In this way, the obtained neutrino mass element $m_{e\mu}$ and the $\nu_\mu \leftrightarrow \bar{\nu}_e$ oscillation probability are not suppressed.

We would like to emphasize the importance of the $U(1)_d$ symmetry in this model. Firstly, without it, the same fields 
may create
the Weinberg operator at tree or one-loop level, spoiling the arguments above. For example, {if $U(1)_d$ is replaced by ${Z}_2$}, the neutral fermions $N_{L,R}$ could obtain their Majorana masses so that the dominant neutrino masses come from the Ma's one-loop diagram enclosed by $N$ and $\phi$ as in Ref.~\cite{Ma:2006km}, which is a simple realization of the Weinberg operator. Furthermore, the presence of the leptoquark $S$ usually involves 
 baryon number violations, such as proton decays, via the following vertices:
\begin{eqnarray}
\bar{e}_R^c S^\dagger u_R + \bar{L}^c_L S^\dagger \tilde{Q}_L  + \bar{d}_R^c u_R S + \bar{Q}_L^c \tilde{Q}_L S + {\rm H.c.}\,,
\end{eqnarray}  
where $\tilde{Q}_L \equiv i\sigma^2 Q_L$. However, the presence of the $U(1)_d$ symmetry forbids the existence of these vertices, so that the baryon number symmetry is still preserved  in the present model.   A further interesting aspect of this $U(1)_d$ symmetry is that it guarantees the lightest neutral particle as a dark matter candidate, which could be the scalar from the mixing of $\sigma$ and the electromagnetic neutral component of $\phi$ or the Dirac fermion $N$. But it is well beyond the scope of the present paper to discuss in detail dark matter physics and other aspects of this model, which we would like to present elsewhere.

\section{Conclusions}\label{s5}
Neutrino-antineutrino oscillations are one of the generic LNV phenomena. However, the conventional theories, including the seesaw neutrino models, are based on the dimension-5 Weinberg operator and predict that such effects are extremely small due to the great suppression from neutrino masses. In the present paper, we have provided a counterexample to such an expectation, in which the neutrino masses originate from the high-dimensional LNV operator ${\cal O}_7$ in Eq.~(\ref{EqO7}) 
 with sizable couplings only to the first two generations of quarks. In this class of models, the LNV in the $\nu$-$\bar{\nu}$ oscillations occurs at the interaction vertices with nuclear targets, rather than through the insertion of the Majorana neutrino masses, so that it escapes the suppressions in the conventional mechanism. Based on our calculations, 
 the $\nu$-$\bar{\nu}$ oscillation probabilities are only mildly suppressed by a factor of ${\cal O}(10^{-6})$ compared with the neutrino-neutrino oscillations, while the suppression factor for the conventional channels is of ${\cal O}(10^{-16}\sim 10^{-22})$. Moreover, such large oscillation probabilities make it possible to observe the $CP$ asymmetries in the $\nu \leftrightarrow \bar{\nu}$ channels which have been shown to be generic in models with complex Wilson coefficients. 
Clearly, our findings reopen the hope to measure these interesting effects by using the conventional neutrino sources such as reactor and accelerator neutrinos.

There are several other interesting features in this class of models characterized by the high-dimensional operator ${\cal O}_7$. Due to the specific dependence on the charged lepton mass hierarchy, the neutrino mass matrix is predicted to be of normal hierarchy. Also, by fitting the neutrino mass data, the cutoff for ${\cal O}_7$ is found to be $\Lambda \sim 1$~TeV by assuming ${\cal O}(1)$ Wilson coefficients  related to the second-generation quarks. Furthermore, the neutrinoless double beta decays are expected to be large in this kind of models, and have already  imposed stringent limits to the Wilson coefficient $C^{ud}_{ee}$. In particular, the $\nu_e \leftrightarrow \bar{\nu}_e$ mode is restricted to be too small to be tested experimentally. However, other modes do not suffer such strong constraints, and can be still large enough to be of phenomenological interest.

Many aspects of the present scenario are worthwhile to be investigated further. Besides of the $0\nu\beta\beta$ decays and $\nu\to\bar{\nu}$ oscillations studied in this paper, there are other LNV effects which are also expected to be large due to new contributions from ${\cal O}_7$. One kind of the promising processes is the LNV rare meson decays~\cite{deGouvea:2002gf,Littenberg:1991ek,Zuber:2000vy, Dib:2000wm,Helo:2010cw,Cvetic:2010rw, Quintero:2011yh, Wang:2014lda,Milanes:2016rzr,Mandal:2016hpr, Quintero:2016iwi},  such as $D_s^{\pm} \to \mu^{\pm}\mu^{\pm}\pi^{\mp}$, which might be tested by the LHCb experiments. Another kind of interesting observables involves the LNV channels at the LHC, {\it e.g.}, the like-sign lepton signature, $pp \to l^\pm l^\pm jj$~\cite{Keung:1983uu,Dicus:1991fk,Atre:2009rg, Atre:2005eb,Ali:2001gsa,Deppisch:2015qwa}. Finally, our UV-complete model realizing ${\cal O}_7$ as the  dominant LNV operator provides a concrete connection between neutrino and dark matter physics, thus deserving detailed studies.  

%%%%%%%%%%%%%%%%%%%%%%%%%%%%%%%%%%%%%%%%%%%%%%%%%%%%%%%%%%%%%%
%%%%%%%%%%%%%%%%%%%%%%%%%%%%%%%%%%%%%%%%%%%%%%%%%%%%%%%%%%%%%%
%%%%%%%%%%%%%%%%%%%%%%%%%%%%%%%%%%%%%%%%%%%%%%%%%%%%%%%%%%%%%%
\section*{Acknowledgments}
We would like to thank Yue-Liang Wu and Bohdan Grzadkowski for helpful discussions. DH would also like to acknowledge the hospitality of Institute of Theoretical Physics, Chinese Academy of Sciences, where part of work was done. DH is supported by the National Science Centre (Poland) research project, decision  DEC-2014/15/B/ST2/00108. CQG is supported by National Center for Theoretical Sciences and MoST (MOST 104-2112-M-009-020-MY3).

%\section*{References}


\begin{thebibliography}{0}
%%%  Neutrino Oscillations  Exp %%%%%%%%%%%%%%%%%%%%%
\bibitem{Anselmann:1992kc}
  P.~Anselmann {\it et al.}  [GALLEX Collaboration],
  %``Implications of the GALLEX determination of the solar neutrino flux.,''
  Phys.\ Lett.\ B {\bf 285}, 390 (1992).
\bibitem{Fukuda:1998mi}
  Y.~Fukuda {\it et al.}  [Super-Kamiokande Collaboration],
  %``Evidence for oscillation of atmospheric neutrinos,''
  Phys.\ Rev.\ Lett.\  {\bf 81}, 1562 (1998)
  [hep-ex/9807003].
\bibitem{Ahmad:2002jz}
  Q.~R.~Ahmad {\it et al.}  [SNO Collaboration],
  %``Direct evidence for neutrino flavor transformation from neutral current interactions in the Sudbury Neutrino Observatory,''
  Phys.\ Rev.\ Lett.\  {\bf 89}, 011301 (2002)
  [nucl-ex/0204008].
\bibitem{Ahmad:2002ka}
  Q.~R.~Ahmad {\it et al.}  [SNO Collaboration],
  %``Measurement of day and night neutrino energy spectra at SNO and constraints on neutrino mixing parameters,''
  Phys.\ Rev.\ Lett.\  {\bf 89}, 011302 (2002)
  [nucl-ex/0204009].
\bibitem{Ahn:2006zza}
  M.~H.~Ahn {\it et al.}  [K2K Collaboration],
  %``Measurement of Neutrino Oscillation by the K2K Experiment,''
  Phys.\ Rev.\ D {\bf 74}, 072003 (2006)
  [hep-ex/0606032].
\bibitem{Abe:2011sj}
  K.~Abe {\it et al.}  [T2K Collaboration],
  %``Indication of Electron Neutrino Appearance from an Accelerator-produced Off-axis Muon Neutrino Beam,''
  Phys.\ Rev.\ Lett.\  {\bf 107}, 041801 (2011)
  [arXiv:1106.2822 [hep-ex]].
\bibitem{An:2012eh}
  F.~P.~An {\it et al.}  [Daya Bay Collaboration],
  %``Observation of electron-antineutrino disappearance at Daya Bay,''
  Phys.\ Rev.\ Lett.\  {\bf 108}, 171803 (2012)
  [arXiv:1203.1669 [hep-ex]].

%%%%%%%%%%%%%%%% %Type I  SeeSaw %%%%%%%%%%%%%%%%%%%%%%%%%%%
\bibitem{TypeIseesaw1}
%\bibitem{Minkowski:1977sc}
  P.~Minkowski,
  %``Mu $\to$ E Gamma At A Rate Of One Out Of 1-Billion Muon Decays?,''
  Phys.\ Lett.\  B {\bf 67}, 421 (1977).
\bibitem{TypeIseesaw2}
T.~Yanagida, in {\it Proceedings of the Workshop on the Unified Theory and the Baryon Number in
the Universe}, edited by O.~Sawada and A.~Sugamoto (KEK, Tsukuba, 1979), p.~95.
\bibitem{TypeIseesaw3}
M.~Gell-Mann, P.~Ramond, and R.~Slansky, in {\it Supergravity},
edited by P.~van Nieuwenhuizen and D.~Freedman (North-Holland, Amsterdam, 1979), p.~315.
\bibitem{TypeIseesaw4}
S.~L.~Glashow, in {\it Proceedings of the 1979 Cargese Summer Institute on Quarks and Leptons},
edited by M.~Levy {\it et al}. (Plenum Press, New York, 1980), p. 687.
\bibitem{TypeIseesaw5}
%\bibitem{Mohapatra:1979ia}
  R.~N.~Mohapatra and G.~Senjanovic,
  %``Neutrino mass and spontaneous parity nonconservation,''
  Phys.\ Rev.\ Lett.\  {\bf 44}, 912 (1980).

%%%%%%%%%%%%%%%%%% TYPE II SEESAW    %%%%%%%%%%%%%%%%%%%%%%%%%%%%%%%%%%
\bibitem{typeIIseesaw1}
M.~Magg and C.~Wetterich,
  %``Neutrino Mass Problem And Gauge Hierarchy,''
  Phys.\ Lett.\  B {\bf 94}, 61 (1980).
\bibitem{typeIIseesaw2}
   J.~Schechter and J.~W.~F.~Valle,
  %``Neutrino Masses In SU(2) X U(1) Theories,''
  Phys.\ Rev.\  D {\bf 22}, 2227 (1980).
\bibitem{typeIIseesaw3}
   T.~P.~Cheng and L.~F.~Li,
  %``Neutrino Masses, Mixings And Oscillations In SU(2) X U(1) Models Of
  %Electroweak Interactions,''
  Phys.\ Rev.\  D {\bf 22}, 2860 (1980).
\bibitem{typeIIseesaw4}
G.~B.~Gelmini and M.~Roncadelli,
  %``Left-Handed Neutrino Mass Scale And Spontaneously Broken Lepton Number,''
  Phys.\ Lett.\  B {\bf 99}, 411 (1981).
\bibitem{typeIIseesaw5}
G.~Lazarides, Q.~Shafi and C.~Wetterich,
  %``Proton Lifetime And Fermion Masses In An SO(10) Model,''
  Nucl.\ Phys.\  B {\bf 181}, 287 (1981).
\bibitem{typeIIseesaw6}
  R.~N.~Mohapatra and G.~Senjanovic,
  %``Neutrino Masses And Mixings In Gauge Models With Spontaneous Parity
  %Violation,''
  Phys.\ Rev.\  D {\bf 23}, 165 (1981).
\bibitem{typeIIseesaw7}
  J.~Schechter and J.~W.~F.~Valle,
  %``Neutrino Decay And Spontaneous Violation Of Lepton Number,''
  Phys.\ Rev.\  D {\bf 25}, 774 (1982).

%%%%%%%%%%%%%%%%%%%  type III Seesaw %%%%%%%%%%%%%%%%%%%%%%%%%%%%%
\bibitem{Foot:1988aq}
  R.~Foot, H.~Lew, X.~G.~He and G.~C.~Joshi,
  %``Seesaw Neutrino Masses Induced by a Triplet of Leptons,''
  Z.\ Phys.\ C {\bf 44}, 441 (1989).

%%%%%%%%%loop-induced neutrino masses
\bibitem{Zee:1980ai}
  A.~Zee,
  %``A Theory of Lepton Number Violation, Neutrino Majorana Mass, and Oscillation,''
  Phys.\ Lett.\ B {\bf 93}, 389 (1980)
  [Erratum-ibid.\ B {\bf 95}, 461 (1980)].
\bibitem{Zee:1985id}
  A.~Zee,
  %``Quantum Numbers of Majorana Neutrino Masses,''
  Nucl.\ Phys.\ B {\bf 264}, 99 (1986).
  %%CITATION = NUPHA,B264,99;%%
\bibitem{Babu:1988ki}
  K.~S.~Babu,
  %``Model of 'Calculable' Majorana Neutrino Masses,''
  Phys.\ Lett.\ B {\bf 203}, 132 (1988).
  %%CITATION = PHLTA,B203,132;%%
  
%%%%%%%%%%%%%%%%Neutrino mass and dark matter
\bibitem{Krauss:2002px}
  L.~M.~Krauss, S.~Nasri and M.~Trodden,
  %``A Model for neutrino masses and dark matter,''
  Phys.\ Rev.\ D {\bf 67}, 085002 (2003)
  [hep-ph/0210389].
\bibitem{Ma:2006km}
  E.~Ma,
  %``Verifiable radiative seesaw mechanism of neutrino mass and dark matter,''
  Phys.\ Rev.\ D {\bf 73}, 077301 (2006)
  [hep-ph/0601225].
\bibitem{Aoki:2008av}
  M.~Aoki, S.~Kanemura and O.~Seto,
  %``Neutrino mass, Dark Matter and Baryon Asymmetry via TeV-Scale Physics without Fine-Tuning,''
  Phys.\ Rev.\ Lett.\  {\bf 102}, 051805 (2009)
  [arXiv:0807.0361 [hep-ph]].
\bibitem{Gustafsson:2012vj}
  M.~Gustafsson, J.~M.~No and M.~A.~Rivera,
  %``Predictive Model for Radiatively Induced Neutrino Masses and Mixings with Dark Matter,''
  Phys.\ Rev.\ Lett.\  {\bf 110}, no. 21, 211802 (2013)
  [Erratum-ibid.\  {\bf 112}, no. 25, 259902 (2014)]
  [arXiv:1212.4806 [hep-ph]].
  
%%%%%%%%%%% Three-Loop Models
%\cite{Chen:2006vn}
\bibitem{Chen:2006vn} 
  C.~S.~Chen, C.~Q.~Geng and J.~N.~Ng,
  %``Unconventional Neutrino Mass Generation, Neutrinoless Double Beta Decays, and Collider Phenomenology,''
  Phys.\ Rev.\ D {\bf 75}, 053004 (2007)
  doi:10.1103/PhysRevD.75.053004
  [hep-ph/0610118].
  %%CITATION = doi:10.1103/PhysRevD.75.053004;%%
  %32 citations counted in INSPIRE as of 20 Oct 2016
%\cite{Chen:2007dc}
\bibitem{Chen:2007dc} 
  C.~S.~Chen, C.~Q.~Geng, J.~N.~Ng and J.~M.~S.~Wu,
  %``Testing radiative neutrino mass generation at the LHC,''
  JHEP {\bf 0708}, 022 (2007)
  doi:10.1088/1126-6708/2007/08/022
  [arXiv:0706.1964 [hep-ph]].
  %%CITATION = doi:10.1088/1126-6708/2007/08/022;%%
  %44 citations counted in INSPIRE as of 20 Oct 2016
\bibitem{Geng:2014gua}
  C.~Q.~Geng, D.~Huang and L.~H.~Tsai,
  %``Loop-induced Neutrino Masses: A Case Study,''
  Phys.\ Rev.\ D {\bf 90}, no. 11, 113005 (2014)
  [arXiv:1410.7606 [hep-ph]].
\bibitem{Jin:2015cla}
  L.~G.~Jin, R.~Tang and F.~Zhang,
  %``A three-loop radiative neutrino mass model with dark matter,''
  Phys.\ Lett.\ B {\bf 741}, 163 (2015)
  [arXiv:1501.02020 [hep-ph]].
\bibitem{Geng:2015coa}
  C.~Q.~Geng, D.~Huang and L.~H.~Tsai,
  %``Comment on "A three-loop radiative neutrino mass model with dark matter" [Phys. Lett. B 741 (2015) 163],''
  Phys.\ Lett.\ B {\bf 745}, 56 (2015)
  [arXiv:1504.05468 [hep-ph]].
%\cite{Geng:2015qha}
\bibitem{Geng:2015qha} 
  C.~Q.~Geng, D.~Huang, L.~H.~Tsai and Q.~Wang,
  %``Connecting Neutrino Masses and Dark Matter by High-dimensional Lepton Number Violation Operator,''
  JHEP {\bf 1508}, 141 (2015)
  doi:10.1007/JHEP08(2015)141
  [arXiv:1507.03455 [hep-ph]].
  %%CITATION = doi:10.1007/JHEP08(2015)141;%%
  %1 citations counted in INSPIRE as of 19 Oct 2016
%\cite{Geng:2015oga}
\bibitem{Geng:2015oga} 
  C.~Q.~Geng, D.~Huang and L.~H.~Tsai,
  %``CP violations in predictive neutrino mass structures,''
  Eur.\ Phys.\ J.\ C {\bf 75}, no. 11, 557 (2015)
  doi:10.1140/epjc/s10052-015-3779-9
  [arXiv:1508.02180 [hep-ph]].
  %%CITATION = doi:10.1140/epjc/s10052-015-3779-9;%%
  %1 citations counted in INSPIRE as of 19 Oct 2016
  

%%%%%%%%%%%%%%%% review for 0nbb decay %%%%%%%%%%%%%%%%%%
%\cite{Pas:2015eia}
\bibitem{Pas:2015eia} 
For an overview, see H.~P${\rm \ddot{a}}$s and W.~Rodejohann,
  %``Neutrinoless Double Beta Decay,''
  New J.\ Phys.\  {\bf 17}, no. 11, 115010 (2015)
  doi:10.1088/1367-2630/17/11/115010
  [arXiv:1507.00170 [hep-ph]].
  %%CITATION = doi:10.1088/1367-2630/17/11/115010;%%
  %34 citations counted in INSPIRE as of 19 Oct 2016

%%%%%%%%%%%%%%%%  Neutrino-antineutrino Oscillations %%%%%%%%%%%%%%%%%
%\cite{Schechter:1980gk}
\bibitem{Schechter:1980gk} 
  J.~Schechter and J.~W.~F.~Valle,
  %``Neutrino Oscillation Thought Experiment,''
  Phys.\ Rev.\ D {\bf 23}, 1666 (1981).
  doi:10.1103/PhysRevD.23.1666
  %%CITATION = doi:10.1103/PhysRevD.23.1666;%%
  %189 citations counted in INSPIRE as of 19 Oct 2016
%\cite{Li:1981um}
\bibitem{Li:1981um} 
  L.~F.~Li and F.~Wilczek,
  %``Physical Processes Involving Majorana Neutrinos,''
  Phys.\ Rev.\ D {\bf 25}, 143 (1982).
  doi:10.1103/PhysRevD.25.143
  %%CITATION = doi:10.1103/PhysRevD.25.143;%%
  %57 citations counted in INSPIRE as of 19 Oct 2016  
%\cite{Bernabeu:1982vi}
\bibitem{Bernabeu:1982vi} 
  J.~Bernabeu and P.~Pascual,
  %``{CP} Properties of the Leptonic Sector for Majorana Neutrinos,''
  Nucl.\ Phys.\ B {\bf 228}, 21 (1983).
  doi:10.1016/0550-3213(83)90393-0
  %%CITATION = doi:10.1016/0550-3213(83)90393-0;%%
  %64 citations counted in INSPIRE as of 19 Oct 2016
\bibitem{Kayser:1984ge}
  B.~Kayser,
  %``CPT, CP, and c Phases and their Effects in Majorana Particle Processes,''
  Phys.\ Rev.\ D {\bf 30}, 1023 (1984).
\bibitem{Langacker:1998pv}
  P.~Langacker and J.~Wang,
  %``Neutrino anti-neutrino transitions,''
  Phys.\ Rev.\ D {\bf 58}, 093004 (1998)
  [hep-ph/9802383].
\bibitem{deGouvea:2002gf}
  A.~de Gouvea, B.~Kayser and R.~N.~Mohapatra,
  %``Manifest CP violation from Majorana phases,''
  Phys.\ Rev.\ D {\bf 67}, 053004 (2003)
  [hep-ph/0211394].
\bibitem{Xing:2013ty}
  Z.~z.~Xing,
  %``Properties of CP Violation in Neutrino-Antineutrino Oscillations,''
  Phys.\ Rev.\ D {\bf 87}, no. 5, 053019 (2013)
  [arXiv:1301.7654 [hep-ph]].
\bibitem{Xing:2013woa}
  Z.~z.~Xing and Y.~L.~Zhou,
  %``Majorana CP-violating phases in neutrino-antineutrino oscillations and other lepton-number-violating processes,''
  Phys.\ Rev.\ D {\bf 88}, 033002 (2013)
  [arXiv:1305.5718 [hep-ph]].
\bibitem{Zhou:2013eoa}
  Y.~L.~Zhou,
  %``Majorana phases in neutrino-antineutrino oscillations,''
  arXiv:1310.5843 [hep-ph].
\bibitem{Xing:2014yka}
  Z.~z.~Xing and Y.~L.~Zhou,
  %``Geometry of the effective Majorana neutrino mass in the $0\nu\beta\beta$ decay,''
  Chin.\ Phys.\ C {\bf 39}, no. 1, 011001 (2015)
  [arXiv:1404.7001 [hep-ph]].
\bibitem{Xing:2014eia}
  Z.~Z.~Xing,
  %``Majorana phases and neutrino-antineutrino oscillations,''
  Int.\ J.\ Mod.\ Phys.\ A {\bf 29}, 1444003 (2014).


%%%%%%%%%%%%%%  High dimensional LNV operators %%%%%%%%%%%%%%
%\cite{Babu:2001ex}
\bibitem{Babu:2001ex} 
  K.~S.~Babu and C.~N.~Leung,
  %``Classification of effective neutrino mass operators,''
  Nucl.\ Phys.\ B {\bf 619}, 667 (2001)
  doi:10.1016/S0550-3213(01)00504-1
  [hep-ph/0106054].
  %%CITATION = doi:10.1016/S0550-3213(01)00504-1;%%
  %107 citations counted in INSPIRE as of 19 Oct 2016
%\cite{deGouvea:2007qla}   
\bibitem{deGouvea:2007qla} 
  A.~de Gouvea and J.~Jenkins,
  %``A Survey of Lepton Number Violation Via Effective Operators,''
  Phys.\ Rev.\ D {\bf 77}, 013008 (2008)
  doi:10.1103/PhysRevD.77.013008
  [arXiv:0708.1344 [hep-ph]].
  %%CITATION = doi:10.1103/PhysRevD.77.013008;%%
  %100 citations counted in INSPIRE as of 19 Oct 2016
%\cite{Angel:2012ug}
\bibitem{Angel:2012ug} 
  P.~W.~Angel, N.~L.~Rodd and R.~R.~Volkas,
  %``Origin of neutrino masses at the LHC: $\Delta L = 2$ effective operators and their ultraviolet completions,''
  Phys.\ Rev.\ D {\bf 87}, no. 7, 073007 (2013)
  doi:10.1103/PhysRevD.87.073007
  [arXiv:1212.6111 [hep-ph]].
  %%CITATION = doi:10.1103/PhysRevD.87.073007;%%
  %40 citations counted in INSPIRE as of 19 Oct 2016
%\cite{Helo:2015fba}
\bibitem{Helo:2015fba} 
  J.~C.~Helo, M.~Hirsch, T.~Ota and F.~A.~Pereira dos Santos,
  %``Double beta decay and neutrino mass models,''
  JHEP {\bf 1505}, 092 (2015)
  doi:10.1007/JHEP05(2015)092
  [arXiv:1502.05188 [hep-ph]].
  %%CITATION = doi:10.1007/JHEP05(2015)092;%%
  %14 citations counted in INSPIRE as of 19 Oct 2016
  

%%%%%%%%%%%%%%%%%%%%%% O8 %%%%%%%%%%%%%%%%%%%%%%%%%%%%%%
%\cite{Babu:2010vp}
\bibitem{Babu:2010vp} 
  K.~S.~Babu and J.~Julio,
  %``Two-Loop Neutrino Mass Generation through Leptoquarks,''
  Nucl.\ Phys.\ B {\bf 841}, 130 (2010)
  doi:10.1016/j.nuclphysb.2010.07.022
  [arXiv:1006.1092 [hep-ph]].
  %%CITATION = doi:10.1016/j.nuclphysb.2010.07.022;%%
  %47 citations counted in INSPIRE as of 19 Oct 2016
%\cite{Cai:2014kra}
\bibitem{Cai:2014kra} 
  Y.~Cai, J.~D.~Clarke, M.~A.~Schmidt and R.~R.~Volkas,
  %``Testing Radiative Neutrino Mass Models at the LHC,''
  JHEP {\bf 1502}, 161 (2015)
  doi:10.1007/JHEP02(2015)161
  [arXiv:1410.0689 [hep-ph]].
  %%CITATION = doi:10.1007/JHEP02(2015)161;%%
  %10 citations counted in INSPIRE as of 19 Oct 2016
  
%%%%%%%%%%%%%%% zero texture  %%%%%%%%%%%%%%%%%%%%%%
\bibitem{PDG}
  K.~A.~Olive {\it et al.}  [Particle Data Group Collaboration],
  %``Review of Particle Physics,''
  Chin.\ Phys.\ C {\bf 38}, 090001 (2014).

\bibitem{Pascoli:2001by}
  S.~Pascoli, S.~T.~Petcov and L.~Wolfenstein,
  %``Searching for the CP violation associated with Majorana neutrinos,''
  Phys.\ Lett.\ B {\bf 524}, 319 (2002)
  [hep-ph/0110287].
  
%%%%%  PMNS %%%%%%%%%%%%%%%%%%%%%%%
\bibitem{Pontecorvo:1957cp}
  B.~Pontecorvo,
  %``Mesonium and antimesonium,''
  Sov.\ Phys.\ JETP {\bf 6}, 429 (1957)
  [Zh.\ Eksp.\ Teor.\ Fiz.\  {\bf 33}, 549 (1957)].

\bibitem{Maki:1962mu}
  Z.~Maki, M.~Nakagawa and S.~Sakata,
  %``Remarks on the unified model of elementary particles,''
  Prog.\ Theor.\ Phys.\  {\bf 28}, 870 (1962).

%%%%%%%%%%%%%%% Large 0nbb decay models %%%%%%%%%%%%%%%
%\cite{delAguila:2011gr}
\bibitem{delAguila:2011gr} 
  F.~del Aguila, A.~Aparici, S.~Bhattacharya, A.~Santamaria and J.~Wudka,
  %``A realistic model of neutrino masses with a large neutrinoless double beta decay rate,''
  JHEP {\bf 1205}, 133 (2012)
  doi:10.1007/JHEP05(2012)133
  [arXiv:1111.6960 [hep-ph]].
  %%CITATION = doi:10.1007/JHEP05(2012)133;%%
  %22 citations counted in INSPIRE as of 20 Oct 2016
%\cite{delAguila:2012nu}
\bibitem{delAguila:2012nu} 
  F.~del Aguila, A.~Aparici, S.~Bhattacharya, A.~Santamaria and J.~Wudka,
  %``Effective Lagrangian approach to neutrinoless double beta decay and neutrino masses,''
  JHEP {\bf 1206}, 146 (2012)
  doi:10.1007/JHEP06(2012)146
  [arXiv:1204.5986 [hep-ph]].
  %%CITATION = doi:10.1007/JHEP06(2012)146;%%
  %31 citations counted in INSPIRE as of 20 Oct 2016
 %\cite{Gustafsson:2014vpa}
\bibitem{Gustafsson:2014vpa} 
  M.~Gustafsson, J.~M.~No and M.~A.~Rivera,
  %``Radiative neutrino mass generation linked to neutrino mixing and 0νββ-decay predictions,''
  Phys.\ Rev.\ D {\bf 90}, no. 1, 013012 (2014)
  doi:10.1103/PhysRevD.90.013012
  [arXiv:1402.0515 [hep-ph]].
  %%CITATION = doi:10.1103/PhysRevD.90.013012;%%
  %19 citations counted in INSPIRE as of 16 Dec 2016}
%\cite{King:2014uha}
\bibitem{King:2014uha} 
  S.~F.~King, A.~Merle and L.~Panizzi,
  %``Effective theory of a doubly charged singlet scalar: complementarity of neutrino physics and the LHC,''
  JHEP {\bf 1411}, 124 (2014)
  doi:10.1007/JHEP11(2014)124
  [arXiv:1406.4137 [hep-ph]].
  %%CITATION = doi:10.1007/JHEP11(2014)124;%%
  %29 citations counted in INSPIRE as of 20 Oct 2016

%%%%%%%%%%%%%%%  0nbb decay %%%%%%%%%%%%%%%%%%%%%%%%
%\cite{Doi:1985dx}
\bibitem{Doi:1985dx}
 M.~Doi, T.~Kotani and E.~Takasugi,
  %``Double beta Decay and Majorana Neutrino,''
  Prog.\ Theor.\ Phys.\ Suppl.\  {\bf 83}, 1 (1985).
  doi:10.1143/PTPS.83.1
  %%CITATION = doi:10.1143/PTPS.83.1;%%
  %603 citations counted in INSPIRE as of 05 Oct 2016

%\cite{Muto:1989cd}
\bibitem{Muto:1989cd}
  K.~Muto, E.~Bender and H.~V.~Klapdor,
  %``Nuclear Structure Effects on the Neutrinoless Double Beta Decay,''
  Z.\ Phys.\ A {\bf 334}, 187 (1989).
  %%CITATION = ZEPYA,A334,187;%%
  %137 citations counted in INSPIRE as of 05 Oct 2016

%\cite{Hirsch:1996qw}
\bibitem{Hirsch:1996qw}
  M.~Hirsch, H.~V.~Klapdor-Kleingrothaus and O.~Panella,
  %``Double beta decay in left-right symmetric models,''
  Phys.\ Lett.\ B {\bf 374}, 7 (1996)
  doi:10.1016/0370-2693(96)00185-2
  [hep-ph/9602306].
  %%CITATION = doi:10.1016/0370-2693(96)00185-2;%%
  %124 citations counted in INSPIRE as of 05 Oct 2016

%%%%%%%%%%%%%%%%0nbb   exp.
\bibitem{Agostini:2013mzu}
  M.~Agostini {\it et al.}  [GERDA Collaboration],
  %``Results on Neutrinoless Double-$\beta$ Decay of $^{76}$Ge from Phase I of the GERDA Experiment,''
  Phys.\ Rev.\ Lett.\  {\bf 111}, no. 12, 122503 (2013)
  [arXiv:1307.4720 [nucl-ex]].
\bibitem{KamLANDZen}
  A.~Gando {\it et al.}  [KamLAND-Zen Collaboration],
  %``Measurement of the double-\beta decay half-life of ^{136}Xe with the KamLAND-Zen experiment,''
  Phys.\ Rev.\ C {\bf 85}, 045504 (2012)
  [arXiv:1201.4664 [hep-ex]].
\bibitem{Gando:2012zm}
  A.~Gando {\it et al.}  [KamLAND-Zen Collaboration],
  %``Limit on Neutrinoless $\beta\beta$ Decay of Xe-136 from the First Phase of KamLAND-Zen and Comparison with the Positive Claim in Ge-76,''
  Phys.\ Rev.\ Lett.\  {\bf 110}, no. 6, 062502 (2013)
  [arXiv:1211.3863 [hep-ex]].
\bibitem{Argyriades:2008pr}
  J.~Argyriades {\it et al.}  [NEMO Collaboration],
  %``Measurement of the Double Beta Decay Half-life of Nd-150 and Search for Neutrinoless Decay Modes with the NEMO-3 Detector,''
  Phys.\ Rev.\ C {\bf 80}, 032501 (2009)
  [arXiv:0810.0248 [hep-ex]].
\bibitem{Arnaboldi:2008ds}
  C.~Arnaboldi {\it et al.}  [CUORICINO Collaboration],
  %``Results from a search for the 0 neutrino beta beta-decay of Te-130,''
  Phys.\ Rev.\ C {\bf 78}, 035502 (2008)
  [arXiv:0802.3439 [hep-ex]].
\bibitem{Arnold:2005rz}
  R.~Arnold {\it et al.}  [NEMO Collaboration],
  %``First results of the search of neutrinoless double beta decay with the NEMO 3 detector,''
  Phys.\ Rev.\ Lett.\  {\bf 95}, 182302 (2005)
  [hep-ex/0507083].
\bibitem{Barabash:2010bd}
  A.~S.~Barabash {\it et al.}  [NEMO Collaboration],
  %``Investigation of double beta decay with the NEMO-3 detector,''
  Phys.\ Atom.\ Nucl.\  {\bf 74}, 312 (2011)
  [arXiv:1002.2862 [nucl-ex]].
 

%%%%%%%%%  mu^- - e^+ Conversion  %%%%%%%%%%%%%%%%%%%%%%%%%
%\cite{Geib:2016atx}
\bibitem{Geib:2016atx} 
  T.~Geib, A.~Merle and K.~Zuber,
  %``$\mu^- - e^+$ conversion in upcoming LFV experiments,''
  Phys.\ Lett.\ B {\bf 764}, 157 (2017)
  doi:10.1016/j.physletb.2016.11.029
  [arXiv:1609.09088 [hep-ph]].
  %%CITATION = doi:10.1016/j.physletb.2016.11.029;%%
  %5 citations counted in INSPIRE as of 03 Feb 2017
  
%\cite{Berryman:2016slh}
\bibitem{Berryman:2016slh} 
  J.~M.~Berryman, A.~de Gouvêa, K.~J.~Kelly and A.~Kobach,
  %``On Lepton-Number-Violating Searches for Muon to Positron Conversion,''
  arXiv:1611.00032 [hep-ph].
  %%CITATION = ARXIV:1611.00032;%%
  %2 citations counted in INSPIRE as of 03 Feb 2017
  
 %\cite{Geib:2016daa}
\bibitem{Geib:2016daa} 
  T.~Geib and A.~Merle,
  %``$\boldsymbol{\mu^-}$- $\boldsymbol{e^+}$ Conversion from Short-Range Operators,''
  arXiv:1612.00452 [hep-ph].
  %%CITATION = ARXIV:1612.00452;%%
  %1 citations counted in INSPIRE as of 03 Feb 2017
  
%\cite{Kaulard:1998rb}
\bibitem{Kaulard:1998rb} 
  J.~Kaulard {\it et al.} [SINDRUM II Collaboration],
  %``Improved limit on the branching ratio of mu- --> e+ conversion on titanium,''
  Phys.\ Lett.\ B {\bf 422}, 334 (1998).
  doi:10.1016/S0370-2693(97)01423-8
  %%CITATION = doi:10.1016/S0370-2693(97)01423-8;%%
  %74 citations counted in INSPIRE as of 03 Feb 2017
  
%\cite{Bartoszek:2014mya}
\bibitem{Mu2e} 
  L.~Bartoszek {\it et al.} [Mu2e Collaboration],
  %``Mu2e Technical Design Report,''
  arXiv:1501.05241 [physics.ins-det].
  %%CITATION = ARXIV:1501.05241;%%
  %64 citations counted in INSPIRE as of 03 Feb 2017
  
%\cite{Cui:2009zz}
\bibitem{COMET} 
  Y.~G.~Cui {\it et al.} [COMET Collaboration],
  %``Conceptual design report for experimental search for lepton flavor violating mu- - e- conversion at sensitivity of 10**(-16) with a slow-extracted bunched proton beam (COMET),''
  KEK-2009-10.
  %%CITATION = KEK-2009-10;%%
  %62 citations counted in INSPIRE as of 03 Feb 2017
  
%%%%%%%%%%%%%%%% LNV Meson decay Experiments %%%%%%%%%%%%%%%%%
%\cite{Massri:2016dff}
\bibitem{Massri:2016dff} 
  K.~Massri [NA48/2 Collaboration],
  %``Searches for Lepton Number Violation and resonances in the $K^{\pm} \to \pi\mu\mu$ decays at the NA48/2 experiment,''
  arXiv:1607.04216 [hep-ex].
  %%CITATION = ARXIV:1607.04216;%%
  %4 citations counted in INSPIRE as of 03 Feb 2017
  
%\cite{Lees:2011hb}
\bibitem{Lees:2011hb} 
  J.~P.~Lees {\it et al.} [BaBar Collaboration],
  %``Searches for Rare or Forbidden Semileptonic Charm Decays,''
  Phys.\ Rev.\ D {\bf 84}, 072006 (2011)
  doi:10.1103/PhysRevD.84.072006
  [arXiv:1107.4465 [hep-ex]].
  %%CITATION = doi:10.1103/PhysRevD.84.072006;%%
  %39 citations counted in INSPIRE as of 03 Feb 2017

%\cite{Miyazaki:2012mx}
\bibitem{Miyazaki:2012mx} 
  Y.~Miyazaki {\it et al.} [Belle Collaboration],
  %``Search for Lepton-Flavor-Violating and Lepton-Number-Violating $\tau \to \ell h h^\prime$ Decay Modes,''
  Phys.\ Lett.\ B {\bf 719}, 346 (2013)
  doi:10.1016/j.physletb.2013.01.032
  [arXiv:1206.5595 [hep-ex]].
  %%CITATION = doi:10.1016/j.physletb.2013.01.032;%%
  %38 citations counted in INSPIRE as of 03 Feb 2017
  
%\cite{Seon:2011ni}
%\bibitem{Seon:2011ni} 
 % O.~Seon {\it et al.} [BELLE Collaboration],
  %``Search for Lepton-number-violating $B^+ \to D^- l^+ l^{\prime +}$ Decays,''
%  Phys.\ Rev.\ D {\bf 84}, 071106 (2011)
 % doi:10.1103/PhysRevD.84.071106
 % [arXiv:1107.0642 [hep-ex]].
  %%CITATION = doi:10.1103/PhysRevD.84.071106;%%
  %38 citations counted in INSPIRE as of 03 Feb 2017
  
%%%%%%%%%%%%%%%%% LFV %%%%%%%%%%%%%%%%%%%%%%%%
%\cite{Adam:2013mnn}
\bibitem{Adam:2013mnn} 
  J.~Adam {\it et al.} [MEG Collaboration],
  %``New constraint on the existence of the $\mu^+ \to e^+\gamma$ decay,''
  Phys.\ Rev.\ Lett.\  {\bf 110}, 201801 (2013)
  doi:10.1103/PhysRevLett.110.201801
  [arXiv:1303.0754 [hep-ex]].
  %%CITATION = doi:10.1103/PhysRevLett.110.201801;%%
  %464 citations counted in INSPIRE as of 03 Feb 2017

%\cite{Aubert:2009ag}
\bibitem{Aubert:2009ag} 
  B.~Aubert {\it et al.} [BaBar Collaboration],
  %``Searches for Lepton Flavor Violation in the Decays tau+- ---> e+- gamma and tau+- ---> mu+- gamma,''
  Phys.\ Rev.\ Lett.\  {\bf 104}, 021802 (2010)
  doi:10.1103/PhysRevLett.104.021802
  [arXiv:0908.2381 [hep-ex]].
  %%CITATION = doi:10.1103/PhysRevLett.104.021802;%%
  %299 citations counted in INSPIRE as of 03 Feb 2017

%\cite{Bertl:2006up}
\bibitem{Bertl:2006up} 
  W.~H.~Bertl {\it et al.} [SINDRUM II Collaboration],
  %``A Search for muon to electron conversion in muonic gold,''
  Eur.\ Phys.\ J.\ C {\bf 47}, 337 (2006).
  doi:10.1140/epjc/s2006-02582-x
  %%CITATION = doi:10.1140/epjc/s2006-02582-x;%%
  %288 citations counted in INSPIRE as of 03 Feb 2017

%\cite{Badertscher:1980bt}
\bibitem{Badertscher:1980bt} 
  A.~Badertscher {\it et al.},
  %``New Upper Limits for Muon - Electron Conversion in Sulfur,''
  Lett.\ Nuovo Cim.\  {\bf 28}, 401 (1980).
  doi:10.1007/BF02776193
  %%CITATION = doi:10.1007/BF02776193;%%
  %16 citations counted in INSPIRE as of 03 Feb 2017

%\cite{Dohmen:1993mp}
\bibitem{Dohmen:1993mp} 
  C.~Dohmen {\it et al.} [SINDRUM II Collaboration],
  %``Test of lepton flavor conservation in mu ---> e conversion on titanium,''
  Phys.\ Lett.\ B {\bf 317}, 631 (1993).
  doi:10.1016/0370-2693(93)91383-X
  %%CITATION = doi:10.1016/0370-2693(93)91383-X;%%
  %262 citations counted in INSPIRE as of 03 Feb 2017
  
%\cite{Honecker:1996zf}
\bibitem{Honecker:1996zf} 
  W.~Honecker {\it et al.} [SINDRUM II Collaboration],
  %``Improved limit on the branching ratio of mu ---> e conversion on lead,''
  Phys.\ Rev.\ Lett.\  {\bf 76}, 200 (1996).
  doi:10.1103/PhysRevLett.76.200
  %%CITATION = doi:10.1103/PhysRevLett.76.200;%%
  %93 citations counted in INSPIRE as of 03 Feb 2017
  
%\cite{Bellgardt:1987du}
\bibitem{Bellgardt:1987du} 
  U.~Bellgardt {\it et al.} [SINDRUM Collaboration],
  %``Search for the Decay mu+ ---> e+ e+ e-,''
  Nucl.\ Phys.\ B {\bf 299}, 1 (1988).
  doi:10.1016/0550-3213(88)90462-2
  %%CITATION = doi:10.1016/0550-3213(88)90462-2;%%
  %507 citations counted in INSPIRE as of 03 Feb 2017
  
%\cite{Hayasaka:2010np}
\bibitem{Hayasaka:2010np} 
  K.~Hayasaka {\it et al.},
  %``Search for Lepton Flavor Violating Tau Decays into Three Leptons with 719 Million Produced Tau+Tau- Pairs,''
  Phys.\ Lett.\ B {\bf 687}, 139 (2010)
  doi:10.1016/j.physletb.2010.03.037
  [arXiv:1001.3221 [hep-ex]].
  %%CITATION = doi:10.1016/j.physletb.2010.03.037;%%
  %185 citations counted in INSPIRE as of 03 Feb 2017


%%%%%%%%%%%%%%%%% LNV Meson Decay %%%%%%%%%%%%%%%%%%%%%%%%%%
%\cite{Littenberg:1991ek}
\bibitem{Littenberg:1991ek} 
  L.~S.~Littenberg and R.~E.~Shrock,
  %``Upper bounds on lepton number violating meson decays,''
  Phys.\ Rev.\ Lett.\  {\bf 68}, 443 (1992).
  doi:10.1103/PhysRevLett.68.443
  %%CITATION = doi:10.1103/PhysRevLett.68.443;%%
  %58 citations counted in INSPIRE as of 20 Oct 2016
%\cite{Zuber:2000vy}
\bibitem{Zuber:2000vy} 
  K.~Zuber,
  %``New limits on effective Majorana neutrino masses from rare kaon decays,''
  Phys.\ Lett.\ B {\bf 479}, 33 (2000)
  doi:10.1016/S0370-2693(00)00333-6
  [hep-ph/0003160].
  %%CITATION = doi:10.1016/S0370-2693(00)00333-6;%%
  %35 citations counted in INSPIRE as of 20 Oct 2016
%\cite{Dib:2000wm}
\bibitem{Dib:2000wm} 
  C.~Dib, V.~Gribanov, S.~Kovalenko and I.~Schmidt,
  %``K meson neutrinoless double muon decay as a probe of neutrino masses and mixings,''
  Phys.\ Lett.\ B {\bf 493}, 82 (2000)
  doi:10.1016/S0370-2693(00)01134-5
  [hep-ph/0006277].
  %%CITATION = doi:10.1016/S0370-2693(00)01134-5;%%
  %52 citations counted in INSPIRE as of 20 Oct 2016
%\cite{Helo:2010cw}
\bibitem{Helo:2010cw} 
  J.~C.~Helo, S.~Kovalenko and I.~Schmidt,
  %``Sterile neutrinos in lepton number and lepton flavor violating decays,''
  Nucl.\ Phys.\ B {\bf 853}, 80 (2011)
  doi:10.1016/j.nuclphysb.2011.07.020
  [arXiv:1005.1607 [hep-ph]].
  %%CITATION = doi:10.1016/j.nuclphysb.2011.07.020;%%
  %37 citations counted in INSPIRE as of 20 Oct 2016
%\cite{Cvetic:2010rw}
\bibitem{Cvetic:2010rw} 
  G.~Cvetic, C.~Dib, S.~K.~Kang and C.~S.~Kim,
  %``Probing Majorana neutrinos in rare K and D, ~D_s, B, B_c meson decays,''
  Phys.\ Rev.\ D {\bf 82}, 053010 (2010)
  doi:10.1103/PhysRevD.82.053010
  [arXiv:1005.4282 [hep-ph]].
  %%CITATION = doi:10.1103/PhysRevD.82.053010;%%
  %48 citations counted in INSPIRE as of 20 Oct 2016
%\cite{Quintero:2011yh}
\bibitem{Quintero:2011yh} 
  N.~Quintero, G.~Lopez Castro and D.~Delepine,
  %``Lepton number violation in top quark and neutral B meson decays,''
  Phys.\ Rev.\ D {\bf 84}, 096011 (2011)
  Erratum: [Phys.\ Rev.\ D {\bf 86}, 079905 (2012)]
  doi:10.1103/PhysRevD.86.079905, 10.1103/PhysRevD.84.096011
  [arXiv:1108.6009 [hep-ph]].
  %%CITATION = doi:10.1103/PhysRevD.86.079905, 10.1103/PhysRevD.84.096011;%%
  %30 citations counted in INSPIRE as of 20 Oct 2016
%\cite{Wang:2014lda}
\bibitem{Wang:2014lda} 
  Y.~Wang, S.~S.~Bao, Z.~H.~Li, N.~Zhu and Z.~G.~Si,
  %``Study Majorana Neutrino Contribution to B-meson Semi-leptonic Rare Decays,''
  Phys.\ Lett.\ B {\bf 736}, 428 (2014)
  doi:10.1016/j.physletb.2014.08.006
  [arXiv:1407.2468 [hep-ph]].
  %%CITATION = doi:10.1016/j.physletb.2014.08.006;%%
  %8 citations counted in INSPIRE as of 20 Oct 2016
%\cite{Milanes:2016rzr}
\bibitem{Milanes:2016rzr} 
  D.~Milanes, N.~Quintero and C.~E.~Vera,
  %``Sensitivity to Majorana neutrinos in $\Delta L=2$ decays of $B_c$ meson at LHCb,''
  Phys.\ Rev.\ D {\bf 93}, no. 9, 094026 (2016)
  doi:10.1103/PhysRevD.93.094026
  [arXiv:1604.03177 [hep-ph]].
  %%CITATION = doi:10.1103/PhysRevD.93.094026;%%
  %10 citations counted in INSPIRE as of 20 Oct 2016
%\cite{Mandal:2016hpr}
\bibitem{Mandal:2016hpr} 
  S.~Mandal and N.~Sinha,
  %``Favoured $B_c$ Decay modes to search for a Majorana neutrino,''
  Phys.\ Rev.\ D {\bf 94}, no. 3, 033001 (2016)
  doi:10.1103/PhysRevD.94.033001
  [arXiv:1602.09112 [hep-ph]].
  %%CITATION = doi:10.1103/PhysRevD.94.033001;%%
  %4 citations counted in INSPIRE as of 20 Oct 2016
%\cite{Quintero:2016iwi}
\bibitem{Quintero:2016iwi} 
  N.~Quintero,
  %``Constraints on lepton number violating short-range interactions from $|\Delta L|=2$ processes,''
  arXiv:1606.03477 [hep-ph].
  %%CITATION = ARXIV:1606.03477;%%




%%%%%%%%%%  like-sign lepton signature at LHC %%%%%%%%%%%%%%%%%%%
%\cite{Keung:1983uu}
\bibitem{Keung:1983uu} 
  W.~Y.~Keung and G.~Senjanovic,
  %``Majorana Neutrinos and the Production of the Right-handed Charged Gauge Boson,''
  Phys.\ Rev.\ Lett.\  {\bf 50}, 1427 (1983).
  doi:10.1103/PhysRevLett.50.1427
  %%CITATION = doi:10.1103/PhysRevLett.50.1427;%%
  %315 citations counted in INSPIRE as of 16 Oct 2016
%\cite{Dicus:1991fk}
\bibitem{Dicus:1991fk} 
  D.~A.~Dicus, D.~D.~Karatas and P.~Roy,
  %``Lepton nonconservation at supercollider energies,''
  Phys.\ Rev.\ D {\bf 44}, 2033 (1991).
  doi:10.1103/PhysRevD.44.2033
  %%CITATION = doi:10.1103/PhysRevD.44.2033;%%
  %55 citations counted in INSPIRE as of 16 Oct 2016
%\cite{Atre:2009rg}
\bibitem{Atre:2009rg} 
  A.~Atre, T.~Han, S.~Pascoli and B.~Zhang,
  %``The Search for Heavy Majorana Neutrinos,''
  JHEP {\bf 0905}, 030 (2009)
  doi:10.1088/1126-6708/2009/05/030
  [arXiv:0901.3589 [hep-ph]].
  %%CITATION = doi:10.1088/1126-6708/2009/05/030;%%
  %333 citations counted in INSPIRE as of 20 Oct 2016
%\cite{Atre:2005eb}
\bibitem{Atre:2005eb} 
  A.~Atre, V.~Barger and T.~Han,
  %``Upper bounds on lepton-number violating processes,''
  Phys.\ Rev.\ D {\bf 71}, 113014 (2005)
  doi:10.1103/PhysRevD.71.113014
  [hep-ph/0502163].
  %%CITATION = doi:10.1103/PhysRevD.71.113014;%%
  %43 citations counted in INSPIRE as of 20 Oct 2016
%\cite{Ali:2001gsa}
\bibitem{Ali:2001gsa} 
  A.~Ali, A.~V.~Borisov and N.~B.~Zamorin,
  %``Majorana neutrinos and same sign dilepton production at LHC and in rare meson decays,''
  Eur.\ Phys.\ J.\ C {\bf 21}, 123 (2001)
  doi:10.1007/s100520100702
  [hep-ph/0104123].
  %%CITATION = doi:10.1007/s100520100702;%%
  %74 citations counted in INSPIRE as of 20 Oct 2016
  
%{\color{red}
%\cite{Deppisch:2015qwa}
\bibitem{Deppisch:2015qwa} 
  F.~F.~Deppisch, P.~S.~Bhupal Dev and A.~Pilaftsis,
  %``Neutrinos and Collider Physics,''
  New J.\ Phys.\  {\bf 17}, no. 7, 075019 (2015)
  doi:10.1088/1367-2630/17/7/075019
  [arXiv:1502.06541 [hep-ph]].
  %%CITATION = doi:10.1088/1367-2630/17/7/075019;%%
  %122 citations counted in INSPIRE as of 03 Feb 2017
%}

\end{thebibliography}
\end{document}